\documentclass[aps,prl,twocolumn,amsmath,amssymb,amsfonts,nofootinbib,long,floatfix]{revtex4}
\usepackage{graphicx,epsfig,latexsym,bm}

\newcommand\G{\mbox{G}}

\newcommand\T{\mbox{T}}
\newcommand\f{\mbox{f}}
\newcommand\atto{\mbox{a}}
\newcommand\Hz{\mbox{Hz}}

\newcommand\MPa{\mbox{MPa}}
\newcommand\km{\mbox{km}}
\newcommand\m{\mbox{m}}
\newcommand\cm{\mbox{cm}}
\newcommand\kg{\mbox{kg}}
\newcommand\g{\mbox{g}}
\newcommand\da{\mbox{d}}

\newcommand\JJ{{\bf{J}}}

\newcommand\mm{\boldsymbol \mu}
\newcommand\oo{\boldsymbol \omega}
\newcommand\BB{{\bf{B}}}

\begin{document}

\title{Can the Blackett conjecture directly account for \\ the magnetic fields of celestial bodies and galaxies? \\
And, is a lab-based test for the Blackett conjecture feasible?}

\author{Leonardo Campanelli$^{1}$}
\email{leonardo.s.campanelli@gmail.com}
\affiliation{$^1$All Saints University, 
5145 Steeles Ave., M9L 1R5, Toronto (ON), Canada}
\date{\today}


\begin{abstract}
No.
\end{abstract}




\maketitle


{\it Introduction.} -- The Blackett hypothesis~\cite{Blackett} is a
conjecture according to which any neutral massive rotating macroscopic body should possess
a magnetic moment $\mm$ proportional to its angular momentum $\JJ$ according to
\begin{equation}
\label{1}
\mm = \beta \, \frac{\sqrt{G}}{2c} \, \JJ.
\end{equation}
Here, $G$ is the Newton constant, $c$ is the speed of light, and $\beta$ is a free
dimensionless constant of order unity.
(In this paper, we use Gaussian-cgs units.)
This effect would have his origin in a putative fundamental unified theory of
gravitation and electromagnetism, in which the ``gravitational magnetism'' would emerge.

The plausibility of the Blackett hypothesis reposes entirely
on empirical ``evidences''. Indeed, Blackett observed in his 1947-paper~\cite{Blackett}
that the magnetic field calculated from Eq.~(\ref{1}) agrees with its observed value
for the Earth, Sun, and 78 Virginis (a spectral type B2 star).
About 32 years later, Sirag~\cite{Sirag} tested the Blackett conjecture
using new available data for Mercury, Venus, Jupiter, Saturn, the Moon, and the pulsar Her X-1.
Equation~(\ref{1}), once again, seemed to be relatively successful in explaining the
observed magnetic fields of celestial bodies.

It is important to stress two points. First, at those times there was no satisfactory
explanation for the existence of
the earth magnetic field and in general of magnetic fields of planets and stars.
Second, although one could expect a correlation between the angular momentum and the
magnetic field of a rotating magnetized body, the impressive feature is that such a
(linear) correlation extended over 15 orders of magnitude in
both $J$ and $\mu$.
Therefore, the Blackett hypothesis, renewed by Sirag,
was perhaps a legitimate tentative to give a theoretical explanation for the
magnetization of so vastly different celestial bodies.

More recently, Opher and Wichoski~\cite{OpherGal} 
have applied the Blackett conjecture to the study of galactic magnetic fields.
Their results suggest that the Blackett effect directly accounts for the
magnetization of galaxies if the Blackett constant $\beta$ is in the range
$10^{-2} \lesssim \beta \lesssim 10^{-1}$.
Jimenez and Maroto~\cite{Jimenez}, on the other hand, have shown that the Blackett hypothesis naturally
emerges in an electromagnetic theory that includes nonminimal couplings to the spacetime curvature.
These analyses seem, once again, not to rule out the Blackett hypothesis.
Recently enough, instead, Barrow and Gibbons~\cite{Barrow2017} have somehow ``relaxed''
the Blackett conjecture by suggesting that
the Blackett's constant is bounded above by a number of order unity, and have verified
their conjecture for (classical) charged rotating black holes in theories where
the exact solution is known.

{\it Limit on Blackett's constant.} -- Planets and satellites of the solar system
are neutral rotating systems which, according to the Blackett conjecture, should be magnetized, and indeed
they are, as revealed by the data of a number of spacecrafts~\cite{Planets}.
Approximating such systems as spheres of radius $R$,
the average magnetic field $\BB$ inside (and on the surface) is proportional to the magnetization,
$\BB = 2\mm/R^3$~\cite{Landau}. Outside the systems, the magnetic field is that of a magnetic dipole with
magnetic moment $\mm$. The angular momentum can be written as
$J =  2\pi I/P$, where $P$ is the intrinsic rotational period, $I=\frac25 k M R^2$
is the moment of inertia, $M$ the mass, and $k$ is the moment-of-inertia parameter
(which for an homogeneous and perfectly spherical object is equal to 1).

A strong constraint on $\beta$ is given by
the non-observation of a dipolar magnetic field of Mars
(yet a residual crustal magnetization has been detected, which seems to point towards an extinct dynamo action).
Using the upper limit on the Martian magnetic dipole moment, $\mu \lesssim 2 \times 10^{20} \G \, \cm^3$~\cite{Mars},
we find
\begin{equation}
\label{0}
\beta \lesssim 2 \times 10^{-5},
\end{equation}
where we used $M = 6.4\times 10^{23} \kg$,  $R = 3390 \km$,  $k=0.925$, and $P = 1.03 \da$~\cite{Planets}.
To our knowledge, this is the strongest constraint on the Blackett's constant.
(In the model of Jimenez and Maroto, the model-dependent limit on the Blackett's
constant comes from the constraints on the parameterized post-Newtonian parameters and turns to be
of order of $\beta \lesssim 10^{-4}$~\cite{Jimenez}.)

With such a low value for the Blackett's constant, planetary magnetic fields
and magnetic fields in stars and galaxies cannot be directly explained by the
Blackett conjecture (the magnetic field produced by the
Blackett effect could act, eventually and at most, as a ``seed'' for those fields).
Moreover, the above limit on $\beta$ makes not feasible,
at the present time, a direct lab-based test of the Blackett conjecture, as we show below.

{\it Barnett effect vs. Blackett effect.} -- It is well known that any (neutral) body rotating at an
angular velocity $\oo$ acquires a magnetic dipole moment.
This effect of ``magnetization by rotation'' is known as Barnett effect~\cite{Barnett}.
For a homogeneous diamagnetic or paramagnetic solid occupying a volume $V$,
the magnetic dipole $\mm$ is~\cite{Landau}
\begin{equation}
\label{Barnett}
\mm = \frac{2m_e c}{e} \, \chi \, g^{-1} V \oo,
\end{equation}
where $m_e$ and $e$ are the mass and electric charge of the electron, $\chi$ is the
volume susceptibility, and $g$ is the gyroscopic $g$-factor.

For a sphere of radius $R$ (the main results do not change if we consider different shapes),
the ratio between the magnetic moment given by the Blackett conjecture and the one given
by the Barnett effect is then
%
%
\begin{equation}
\label{ratio}
\frac{\mu \, \mbox{(\small Blackett)}}{\mu \, \mbox{(\small Barnett)}}
\sim 10^{-3} \! \left( \frac{\beta}{10^{-5}} \right) \!\!
\left( \frac{10^{-6}\cm^3/\g}{\chi_m} \right) \!\! \left( \frac{R}{1 \m} \right)^{\!2} \! ,
\end{equation}
where $\chi_m = \chi/\rho$ is the mass susceptibility and $\rho$ the density.
To our knowledge, there are not known solid materials with mass susceptibility below
$10^{-6}\cm^3/\g$. Equation~(\ref{ratio}), then, shows that the Blackett effect is always
subdominant with respect to the Barnett one for lab-scale objects. Our conclusion is that,
at the present time, the Blackett effect cannot be tested in a laboratory.

The Blackett effect could eventually be tested if a material with a mass
susceptibility as low as $10^{-9}\cm^3/\g$ were synthesized.
This is in principle possible if one combines two or more inert materials with
different magnetic properties.
According to the Wiedemann's additivity law~\cite{Wiedemann},
the mass susceptibility of a mixture of its paramagnetic and diamagnetic components would be
$\chi_m = \sum_i m_i \chi_m^{(i)}/\sum_i m_i$, where $m_i$ and $\chi_m^{(i)}$ are the mass
and mass susceptibility of the component $i$.
Thus, an appropriate choice of the mass percentage of each constituent in powder form would give
the possibility of obtaining a material with magnetic susceptibility as low as desired.
The resulting powder could be then sintered and made into a solid.
It is worth noticing that such a procedure has been already applied by Khatiwada et al. to produce a
solid material with very low (volume) susceptibility composed by tungsten and bismuth~\cite{Khatiwada}.
However, even if the resulting solid pellets were compact enough to stay together they were delicate.
According to Khatiwada et al., the pressing procedure could be further enhanced by using higher pressures
and temperatures to produce strong solids.
Even if this were possible, however, the resulting solid material should have a
sufficiently large volume, and be dense and strong enough in order to produce a
detectable magnetic field once it is put into rotational motion, as we discuss below.

{\it Blackett-type experiment.} -- Let us consider a homogeneous rotating sphere made of
a hypothetical material whose mass susceptibility is such that the Blackett effect is
dominant with respect to the Barnett one.
The maximum safe angular speed $\omega$ can be found as follows.
The stress tensor in spherical coordinates $r, \theta, \phi$ can be written as
$\sigma_{ij} = c_{ij}(\nu,\theta,r) \rho \, \omega^2 R^2$, where
$c_{ij}(\nu,\theta,r)$ is a dimensionless tensor, $\nu$ is the Poisson's ratio~\cite{Landau2},
and $i,j = r, \theta, \phi$.
Here, $\sigma_{rr}$ is the radial stress, $\sigma_{r\theta}$ is the shear stress, and
$\sigma_{\theta \theta}$ and $\sigma_{\phi \phi}$ are the angular normal stresses (all the other components
of the stress tensor are zero by symmetry). Using the results of~\cite{stress} we find
that, for given density, angular speed, and radius,
the maximum stress corresponds to the angular normal stresses and in
particular $\max_{\theta,r} c_{\theta \theta} = \max_{\theta,r} c_{\phi \phi} = c_\nu$,
where $c_\nu = (5 \nu^2-\nu-12)/(25\nu^2+10\nu-35)$.
(Here, we assumed $0 \leq \nu \leq 1/2$, which is certainly true for all metals and
known alloys. Theoretically, the Poisson's ratio is in the range $-1 \leq \nu \leq 1/2$~\cite{Landau2}.)
The function $c_\nu$ is an increasing function of $\nu$ such that $c_0 = 12/35 \simeq 0.34$
and $c_{1/2} = 9/19 \simeq 0.47$.

The stress generated by rotation must be smaller that the ultimate tensile stress $\sigma_{\rm max}$,
$|\sigma_{ij}| < \sigma_{\rm max}$. 
This, in turn, determines the maximum possible value for the
angular speed, 
$\omega_{\rm max} = (\sigma_{\rm max}/c_\nu \rho)^{1/2}/R$. Inserting this value of $\omega$
in Eq.~(\ref{1}), we find the maximum magnetic field that can be generated by a rotating
sphere, 
%
%
\begin{equation}
\label{Bmax}
B_{\rm max} \sim 10^{-13} \! \left( \frac{\beta}{10^{-5}} \right) \!\!
\left( \frac{\sigma_{\rm max}}{1 \MPa} \right)^{\!\!1/2} \! \left( \frac{\rho}{1 \g/\cm^3} \right)^{\!\!1/2}
\!\! \left( \frac{R}{1 \m} \right) \! \G.
\end{equation}
For a given radius $R$, then, the maximum magnetic field is large for materials with
high density and ultimate tensile stress,
such as metals and alloys (the dependence of $B_{\rm max}$ and $\omega_{\rm max}$ on
the Poisson's ratio is very week).

In order to detect $B_{\rm max}$, or to put a limit on the Blackett's constant
more stringent than the one in Eq.~(\ref{1}), the hypothetical material 
must have a sufficiently high ultimate tensile stress and density. Indeed,
taking $\beta = 10^{-5}$, a 2-meter sphere ($R = 1\m$) would produce a maximal
magnetic field of order of
$B_{\rm max} \sim 10^{-17} (\sigma_{\rm max}/1 \MPa)^{1/2} (\rho/1 \g \, \cm^{-3})^{1/2}$ \T.
The most sensitive magnetometers are SQUID magnetometers, with maximum
sensitivities of order of $1 \f\T/\sqrt{\Hz}$~\cite{SQUID}, and
SERF magnetometers, with maximum sensitivities of about $0.2 \f\T/\sqrt{\Hz}$~\cite{SERF}.
Even taking the maximum theoretical sensitivity of a SERF magnetometer, estimated to be $2\atto \T$~\cite{SERF2},
the hypothetical material must satisfy the mechanical condition
$(\sigma_{\rm max}/1 \MPa)^{1/2} (\rho/1 \g \, \cm^{-3})^{1/2} \gtrsim 0.1$
in order to be of any relevance.
(As a reference, a relatively low-density material with relatively low ultimate tensile stress is the
``normal strength Portland'' cement concrete for which $\rho \simeq 2.3 \g/\cm^3$ and
$\sigma_{\rm max} \simeq 3.5 \MPa$~\cite{concrete},
while a very strong and very dense material
is tungsten for which $\sigma_{\rm max} \simeq 1510 \MPa$ and $\rho \simeq 19.25 \g/\cm^3$~\cite{Tungsten}.)

{\it Conclusions.} -- The Blackett effect is a hypothetical effect
consisting in the magnetization by rotation of a rigid neutral body
that should emerge from a unified theory of gravitation and electromagnetism.

We have derived a stringent constraint on the Blackett's constant, the
dimensionless constant of proportionality between the magnetization and the angular momentum
of a body, by using the data on the dipolar magnetic field of Mars.
This constraint excludes the possibility that the Blackett effect could directly account for
planetary, stellar, and galactic magnetic fields.

We have also pointed out that the Blackett effect is similar but subdominant for lab-scale objects
with respect to the well-known and experimentally tested Barnett effect,
according to which any rotating object acquires
a magnetic moment proportional to its angular velocity.
The Blackett effect, then, cannot be tested in a laboratory.

\vspace*{0.2cm}

We would like to thank M. Giannotti for useful discussions.


\end{document}